\documentclass{appolb}

\Preprinttrue

\usepackage{epsfig}
\usepackage{url}
\usepackage{graphics}
\usepackage{wrapfig}
\newcommand{\GeV}{\;\mathrm{GeV}}
\newcommand{\as}{\alpha_s}
\newcommand{\order}[1]{\mathcal{O}\left(#1\right)}
\newcommand{\TeV}{\,\mathrm{TeV}}
\newcommand{\h}{\mathrm{h}}
\newcommand{\UE}{\mathrm{UE}}

\begin{document}
\title{Recent progress in defining and understanding jets%
  \thanks{Presented at the 37th International Symposium on
    Multiparticle Dynamics, Berkeley, USA, August 2007}%
}
\author{Gavin P. Salam
\address{LPTHE, CNRS UMR 7589; Universit\'e Pierre et Marie Curie (Paris VI);
Universit\'e Denis Diderot (Paris VII), 75252 Paris Cedex 05, France\vspace{-1.0em}}
%\and
%the Name(s) of other Author(s)
%\address{and their affiliation}
}
\maketitle
\begin{abstract}
  This talk reviews some key developments that have taken place in
  hadron-collider jet finding over the past couple of years,
  including: technical advances such as the complete formulation of an
  infrared safe seedless cone algorithm and fast computational
  approaches to sequential recombination jet finders like the $k_t$
  algorithm, together with universal methods for subtracting pileup;
  progress in understanding the sensitivity of jet algorithms to the
  underlying event and hadronisation; and work that exploits our
  knowledge of QCD divergences to better define and predict
  heavy-flavour jet cross sections.
\end{abstract}
%\PACS{PACS numbers come here}
  
%----------------------------------------------------------------------
\section{Introduction}

Jet algorithms provide a way of projecting away the multiparticle
dynamics of an event so as to leave a simple quasi-partonic picture of
the underlying hard scattering. 
This projection is however fundamentally ambiguous,
reflecting the divergent and quantum mechanical nature of QCD.
Consequently, jet physics is a rich subject.

Key developments in the history of jet finding have often been spurred
by advances in experimental sophistication, and in this vein, the
upcoming startup of the LHC provides a motivation for reexamining the
technology at our disposal. 

To appreciate what changes at LHC, consider the physics scales and
processes at play: in addition to having the electroweak ($\sim
100\GeV$) and hadronisation ($0.5\GeV$) scales familiar at LEP and
HERA, and an underlying event ($\sim 10 \GeV$) 2--4 times larger than
the Tevatron's, the LHC will routinely probe multi-jet events of
unprecedented complexity (think $t\bar t H\to 8\,\mathrm{jets}$), it
will suffer from huge pileup ($\sim 100 \GeV$ of $p_t$ per unit
rapidity),
and it may well discover new-particle cascades that mix the TeV scale
and electroweak scales.
That's vastly more to disentangle than ever before.
To add to that, a key technical issue is posed by the number of
particles: up to $\sim 4000$ per event, two orders of magnitude larger than at
LEP ($\sim 50$), and an order larger than at Tevatron.

A programme of work to bring jet-finding up-to-date for the LHC age
has begun over the past couple of years and involves three main
phases:
1) develop a core set of theoretically solid and experimentally
practical jet algorithms (\ie the 1990 Snowmass
accord~\cite{Snowmass});
2) quantify, where possible analytically, how jet algorithms respond
to various non-perturbative and perturbative QCD effects;
3) use the resulting understanding to guide development of more
sophisticated tools. As described below, phase 1 is nearing
completion, and progress is being made on the remaining parts.

\section{Core tools}

The 1990 Snowmass accord~\cite{Snowmass} for the Tevatron advocated
the use of jet algorithms that were simple to use, both theoretically
and experimentally, well-defined and finite at all orders of
perturbation theory, and relatively insensitive to hadronisation.
However this accord has never fully been satisfied in Tevatron jet
finding.

\textbf{Cone algorithms} provide a top-down form a jet identification,
and are mostly based on the idea of a \emph{stable cone}, one whose
direction coincides with that of the summed momenta of the contained
particles.  They are widespread at $pp$ colliders, motivated on
the grounds that soft and collinear radiation leaves stable cones
unchanged, and a feature often quoted as being one of their main
experimental advantages is their simple conical shape.

Cone algorithms have long been plagued by infrared and collinear (IRC)
safety issues. Old iterative cones with split-merge procedures are
unreliable from order $\as^3$ (or $\alpha_{EW}\as^2$) onwards,
Tevatron run~II ``midpoint'' types cones from order $\as^4$ (or
$\alpha_{EW}\as^3$). Unreliable at order $\as^n$ means here that they
diverge when calculated at $\as^{n+1}$ --- regulating that divergence
around $\Lambda_{QCD}$ introduces a term $\sim\as^{n+1} \ln
p_t/\Lambda_{QCD}$. This is the same size as the $\as^{n}$ term
(recall $\as \sim 1/\ln (p_t/\Lambda_{QCD})$), \ie any effort%
\footnote{Between $50$ and $100$ people working over ten years, \ie
  $\sim \$ 50$ million.} %
that went into the $\as^{n}$ precision is swamped by the
near-divergent uncalculated higher orders.

The practical relevance of the IRC safety issue has been repeatedly
questioned, it being noted for example that in the most widely studied
of jet-observables, the inclusive-jet spectrum, the ``real'' effect of
IR unsafety is seen to be $1\%$.
This figure however holds only for this observable: leading order for
the jet-spectrum is $\as^2$, the midpoint cone's unreliability starts
at $\as^4$, and for $\as\simeq 0.1$ the ratio is $1\%$. At LHC, many
interesting processes \emph{start} at $\as^4$ or higher, and then even
leading order can be unreliable, with up to $50\%$ effects,
depending on the cuts.

\begin{figure}
  \centering
  \begin{minipage}[c]{0.32\linewidth}
    \includegraphics[height=\textwidth,angle=270]{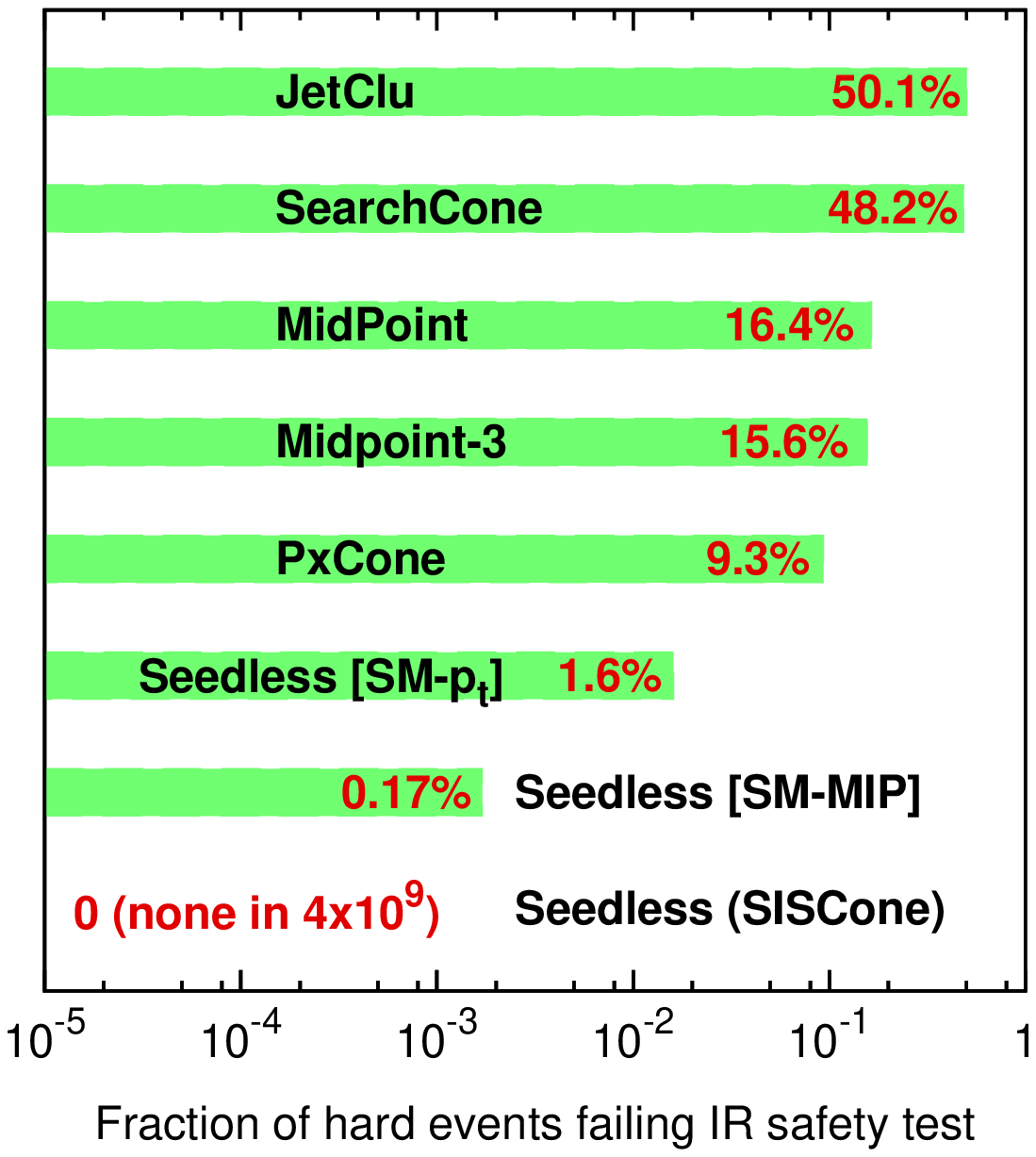}%
  \end{minipage}%
  \begin{minipage}[c]{0.67\linewidth}
    \includegraphics[width=\textwidth]{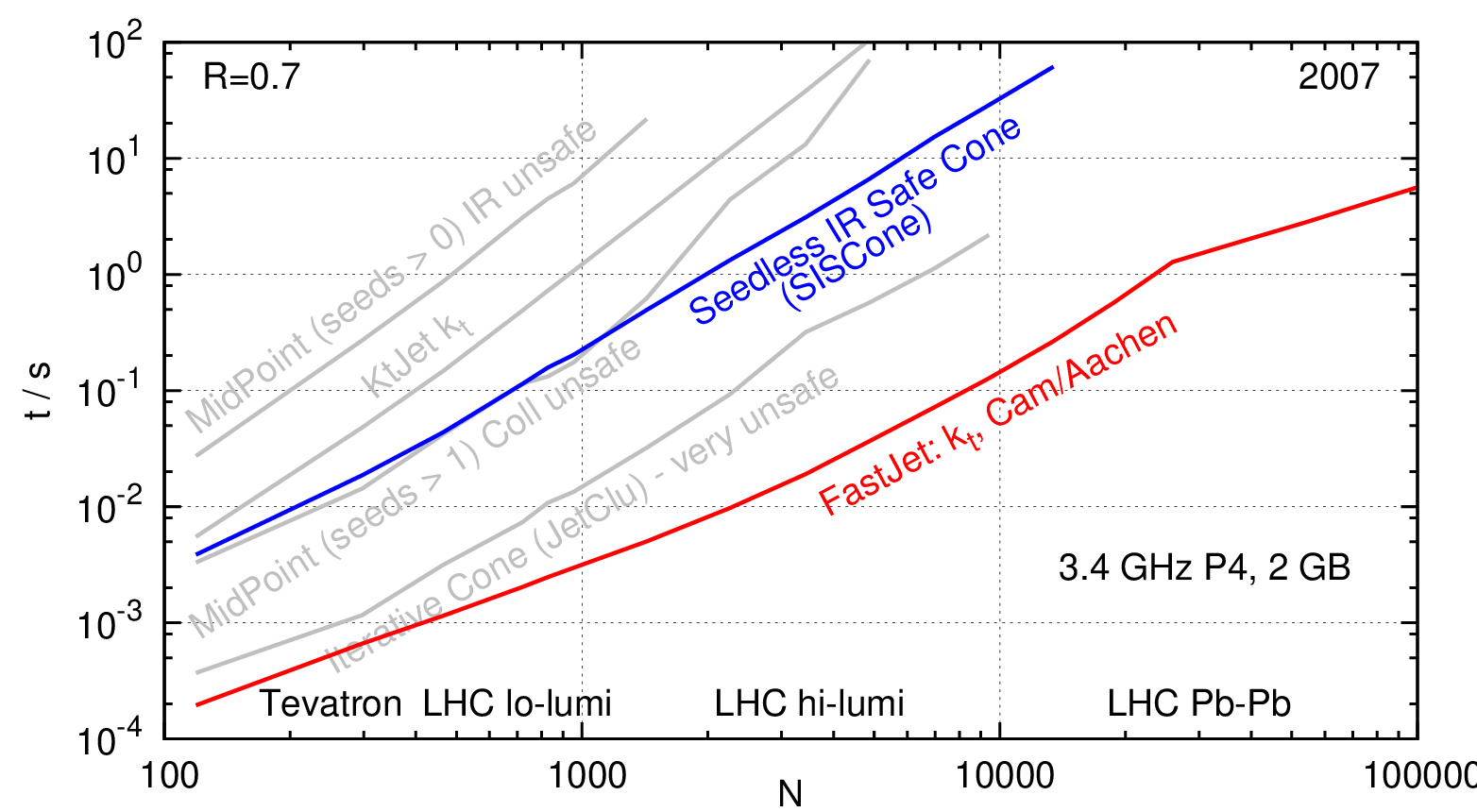}
  \end{minipage}
  \caption{Left: IR safety failure rate for a range of jet algorithms
    in artificial events with between 2 and 10 hard particles (for
    details, see \cite{SISCone}).  Right: speeds of various
    algorithms as a function of the particle multiplicity $N$.}
  \label{fig:speeds}
\end{figure}

The design of an IRC safe cone algorithm starts with the observation
that you should find \emph{all} stable cones~\cite{KOS}.
Ref.~\cite{RunII-jet-physics} showed how, for a handful of particles
(in $N2^N$ time, \ie $10^{17}$ years for $N=100$).  Recently,
Ref.~\cite{SISCone} reduced that to a more manageable $N^2 \ln N$. The
trick was to recast it as a computational geometry problem, \ie the
identification of all distinct circular enclosures for points in 2D,
and to find a (previously unknown) solution to that.  Together with a
few other minor fixes, this has led to the first ever IRC safe jet
algorithm, SISCone (cf. left plot of fig.~\ref{fig:speeds}).

%----------------------------------------------------------------------
%\section{Technical advances}

%----------------------------------------------------------------------
%\section{Sequential recombination algorithms}
\textbf{Sequential recombination algorithms} (SRAs), such as $k_t$
\cite{kt}, take a bottom-up approach to creating jets, successively
merging the closest pair of objects in an event until all are
sufficiently well separated.  They work because of relations between
the distance measures that are used
and the divergences of QCD. Their attractions include their conceptual
simplicity, as well as the hierarchical structure they ascribe to an
event, and they were ubiquitous at LEP and HERA.

There had been two major issues for SRAs in $pp$ collisions: they used
to be slow ($\sim N^3$ time to cluster $N$ particles, \ie 1 minute for
$N=4000$) and the shape of the resulting jets was unknown and
irregular, which complicated pileup subtraction.
Recently the speed issue was solved \cite{FastJet} by observing a
connection with computational geometry problems: \eg the $k_t$
algorithm factorises into a priority queue and the problem of
constructing a nearest-neighbour-graph in 2D and maintaining it under
point changes (solved in \cite{CGALTriang}).
%; C/A reduces to the dynamic 2D
%closest-pair problem. 
Asymptotically, run times are now $N \ln N$, and in practice $\sim
20\,$ms for $N=4000$. That's better even than a fast (but very IR
unsafe) iterative cone algorithm such as CDF's JetClu (cf. right-plot
of fig.~\ref{fig:speeds}).

\begin{figure}
  \centering
  \begin{minipage}[c]{0.4\linewidth}
    \includegraphics[width=1.1\textwidth,height=1.05\textwidth]{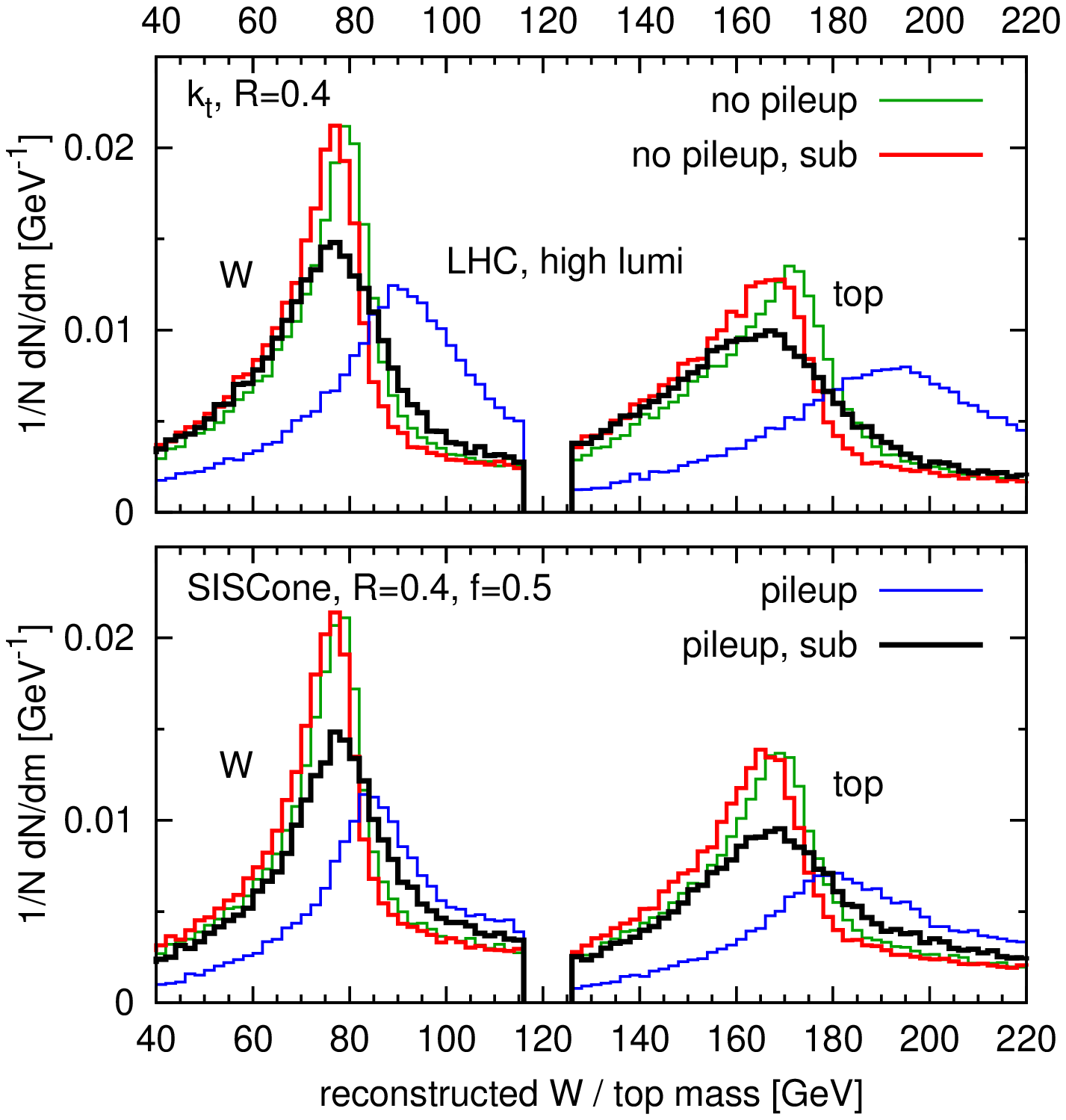}\hfill
  \end{minipage}\hfill
  \begin{minipage}[c]{0.55\linewidth}
    \includegraphics[width=\textwidth]{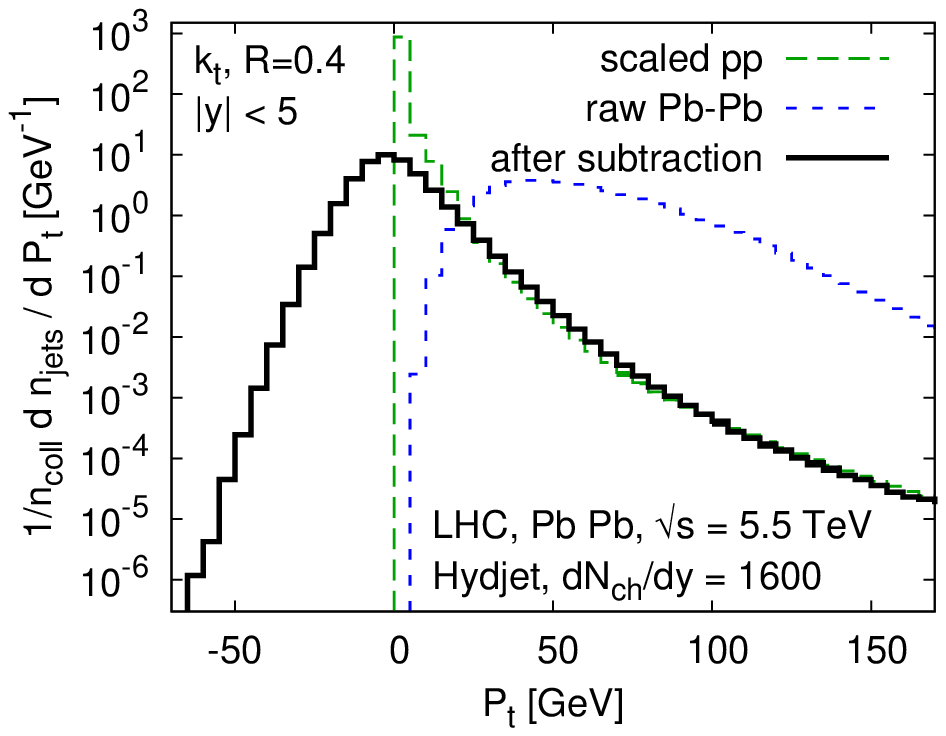}
  \end{minipage}%
  \caption{Left: hadronic $W$ and top mass peaks with pileup
    subtraction in high-luminosity semi-leptonic $t \bar t$ events at
    LHC.  Right: inclusive jet spectrum in LHC PbPb events before and
    after background subtraction. Adapted from \cite{Subtraction},
    simulations used Pythia \cite{Pythia} and Hydjet
    \cite{hydjet}.\vspace{-1.5em}}
  \label{fig:subtraction}
\end{figure}

The problem of the unknown shape of SRA jets has also been solved, by
the simple expedient of adding very many infinitely soft ``ghost''
particles \cite{Area}. These serve to fill in all empty space in the
event and so give a well-defined boundary and total area to each jet.
Subtracting a correction proportional to that area works rather well
for removing pileup \cite{Subtraction} and can even be applied to the
extremely noisy environment of LHC Pb$\,$Pb collisions
(fig.~\ref{fig:subtraction}).
This progress, together with the the recent successful measurement of
the inclusive jet spectrum by CDF with the $k_t$ algorithm
\cite{CDFkt}, means that all objections raised in the past about SRAs
are now essentially resolved.

%----------------------------------------------------------------------
\section{Understanding and improving jet algorithms}

\begin{wrapfigure}{R}{0pt}
  \includegraphics[width=0.42\textwidth]{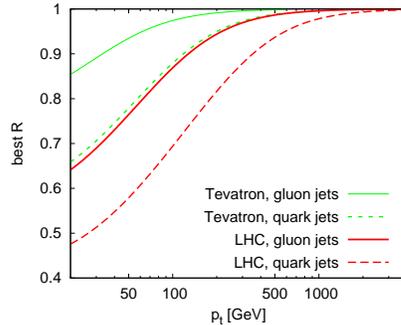}
  \caption{Simple estimate for optimal $R$ as a function of jet $p_t$,
    collider and initiating parton.}
  \label{fig:opt-R}
\end{wrapfigure}
Once you have a set of safe, fast algorithms (all conveniently
packaged in \texttt{FastJet} \cite{FastJet}), you can start trying to
understand their physics behaviour. A simple question, for example, is
that of how \textbf{hadronisation and the underlying event} (UE)
modify a jet's transverse momentum.  The situation is summarised in
table~\ref{tab:summary-of-effects} \cite{DMSHadr}, whose results are
essentially common to all jet algorithms with a radius parameter $R$.
The distinct $R$-dependence for each effect may provide a way of
disentangling them experimentally. It also implies an optimal $R$
(minimising the sum of squares of effects) that varies significantly
with the jet initiator's colour and $p_t$, as illustrated in
fig.~\ref{fig:opt-R}, based on the results in
table~\ref{tab:summary-of-effects}: %
at large $p_t$ perturbative radiation dominates over other
contributions, so one prefers $R\sim 1$, whereas at low $p_t$ the UE
has a significant relative impact and it is advantageous to lower $R$
to limit this, especially at higher energy colliders (since the UE
grows with $s$) and for quark jets (for which perturbative radiation
is weaker). Note that these results for the optimal $R$ are mainly to
be taken as indicative of general trends: a definitive estimate would
go beyond the small-$R$ approximation and take into account the
dispersion of each effect rather than its mean value.

\begin{table}
  \centering\small
  \begin{tabular}{l|c|c|c|c|}
    & \multicolumn{4}{c|}{Jet $\langle \delta p_t \rangle$ given by
      product of dependence on }\\
    & scale & colour factor &  $R$ & $\sqrt{s}$\\\hline
    perturbative radiation & $\sim \frac{\as(p_t)}{\pi} \, p_t$ & $C_i$ & $\ln R +
    \order{1}$ & -- \\[3pt]
    hadronisation   & $\Lambda_{\h}$ & $C_i$ & $-1/R + \order{R}$ &  -- \\
    underlying event  & $\Lambda_{\UE}$ & -- & $R^2/2 + \order{R^4}$ &
    $s^{\omega}$ \\\hline
  \end{tabular}
  \caption{Summary of the main physical effects that contribute
    to the average difference $\langle\delta p_t\rangle$ between the
    transverse momentum of a jet and  its parent
    parton (for small $R$). $\Lambda_{\h}\simeq 0.35-0.4 \GeV$ based
    on $e^+e^-$
    event-shape studies \cite{Dasgupta:2003iq}, $\Lambda_{\UE} s^{\omega}
    \simeq 4\GeV\cdot\left(s/(2\TeV)^2\right)^{0.25}$~\cite{DMSHadr}.}
  \label{tab:summary-of-effects}
\end{table}

While to a first approximation the effects shown in
table~\ref{tab:summary-of-effects} are independent of the specific jet
algorithm, more refined studies, for jet areas \cite{Area}, do
highlight differences between algorithms, but not always as one would
expect. For example, with heavy pileup, the $k_t$ algorithm, often
labelled a vacuum cleaner, actually has an average area quite close to
$\pi R^2$ (essentially because pileup is not vacuum); cone algorithms
are widely assumed to have an area $\pi R^2$, but modern versions with
split-merge steps (\eg SISCone) actually turn out not to be quite
conical, having an area $\sim \pi R^2/2$. This \emph{small} area is
part of the reason why they work well in noisy
environments.\footnote{On the other hand, for hard particles, modern
  cone algorithms like SISCone have a reach that extends somewhat
  beyond $R$ and this can lead to issues in resolving complex multi-jet
  events.} %
This has important implications for strategies that assume an area of
$\pi R^2$ in correcting for pileup with cone-type algorithms.

Perhaps the most striking example to date where a better understanding
of clustering dynamics can lead to improved algorithms concerns
\textbf{jet flavour}. This concept is often taken for granted (over
350 articles' titles contain the words `quark jet' or `gluon jet'),
and it would seem that if one simply sums the flavours of all partons
in a jet one might obtain a well-defined result for the jet-flavour.
This turns out not to be the case, even in algorithms for which the
jet momenta are IRC safe, because the flavour is subject to
contamination by large-angle $g\to q\bar q$ splitting of a soft gluon,
where the $q$ and $\bar q$ then enter separate jets. A simple
modification \cite{Banfi:2006hf} of the distance measure in the $k_t$
algorithm can solve the problem and make the flavour IRC safe.

\begin{wrapfigure}{R}{0pt}
  \includegraphics[height=0.45\textwidth,angle=270]{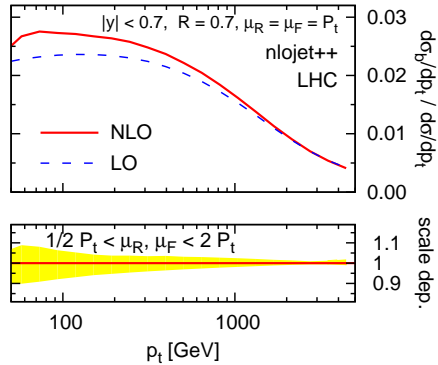}
  \caption{$b$-jet fraction at LHC and scale dependence, with
    flavour-$k_t$ jets~\cite{Banfi:2007gu}.}
  \label{fig:flav-kt}
\end{wrapfigure}
A key advantage of the resulting IRC safe ``flavour-$k_t$'' algorithm
emerges when talking about \emph{heavy}-flavour jet cross sections.
With unsafe definitions, higher orders involve powers of large
logarithms $\ln p_t/m_b$, giving large NLO scale uncertainties.
This is especially the case for current experimental $b$-jet
measurements, in which a jet containing a $b$ and a $\bar b$ is
considered to be a normal $b$-jet.
With a proper, IRC safe definition most of the large logarithms
disappear, and remaining ones can be absorbed into the parton
distribution functions.
The result is a reduction in the theory uncertainty for the inclusive
$b$-jet spectrum from $\sim 40-50\%$ (see e.g.\ \cite{cdf-bjets}) to
the $\sim 10-20\%$ shown in fig.~\ref{fig:flav-kt} \cite{Banfi:2007gu}
(calculated with a modified version of \texttt{nlojet++}
\cite{NLOJET}).  It should be said that while normal IRC safe jet
algorithms involve no particular experimental issues, the
flavour-$k_t$ algorithm does require a jet's flavour to be taken as
the sum of the flavours of the jet's constituents --- \ie
one should be able to distinguish a jet containing a $b$ and $\bar b$
(not a $b$-jet) from one that contains just a single $b$.  This is
challenging experimentally, but if it can be done it will have
significant benefits also in reducing QCD ``$b$''-jet backgrounds
(more than $50\%$ of which come from $b \bar b$ jets) in new-physics
searches. A measurement of the flavour-$k_t$ $b$-jet spectrum would
then provide a powerful cross-check that the separation of $b\bar b$
and $b$ jets is being done effectively.

%----------------------------------------------------------------------
\section{Outlook}

Practical, infrared and collinear safe options now exist for both cone
and sequential recombinations jet algorithms, and are in the process
of being incorporated into the LHC experiments' software frameworks.
If they are widely adopted (not to be taken for granted given the
continued presence also of long-established unsafe options and the
inertia inherent in large organisations), hadron-collider jet-finding
will finally come into accord with the 1990 Snowmass principles, a key
step if the LHC is to benefit from the ongoing and extensive
calculational effort in QCD.

There is more, of course, to jet finding than just practicality and
IRC safety. Most of the thought about jet algorithms, historically,
has been for low-noise, single-scale, quark-jet dominated environments
such as LEP. Work is now being carried out that addresses the
significant novel issues that arise at the LHC. There is considerable
scope for further work, and it is to be hoped that this, together with
open exchanges between theorists and experimenters, will help us make
the best possible use of jets in the rich environment of LHC.\medskip

\textbf{Acknowledgements.} I wish to thank Andrea Banfi, Matteo
Cacciari, Mrinal Dasgupta, Lorenzo Magnea, Gregory Soyez and Giulia
Zanderighi for stimulating collaborations on the topics described
here, as well as G\"unther Dissertori, Steve Ellis, Joey Huston and
Markus Wobisch for numerous interesting discussions.  Work supported
in part by contract ANR-05-JCJC-0046-01.%\vspace{-0.5cm}

\end{document}